\documentstyle[12pt,aaspp4]{article}

\def\c2{[\ion{C}{2}]}
\def\co{$\, ^{\rm 12}$CO$\,$}
\def\g0{$G_{0}$}
\received{22 January 1997}
\accepted{14 July 1997}
\lefthead{Matsuhara et al.}
\righthead{[\ion{C}{2} Observation of High-latitude Molecular Clouds}

\slugcomment{\hfill {\tt DPNU-97-35} \\
\hfill to appear in ApJ Vol.490,  December 1, 1997 issue}

\begin{document}

\title{Observations of \c2 158 micron Line and Far-infrared Continuum Emission
toward the High-latitude Molecular Clouds in Ursa Major}

\author{Hideo Matsuhara, Masahiro Tanaka\altaffilmark{1}, Yoshinori Yonekura, 
and Yasuo Fukui}
\affil{Department of Astrophysics, Nagoya University, Furo-cho, Chikusa-ku,
 Nagoya 464-01, Japan; \\
maruma@toyo.phys.nagoya-u.ac.jp, tanaka@toyo.phys.nagoya-u.ac.jp, 
yonekura@a.phys.nagoya-u.ac.jp, fukui@a.phys.nagoya-u.ac.jp}
\author{Mitsunobu Kawada}
\affil{the Institute of Space and Astronautical Science, 3-1-1 Yoshinodai,
 Sagamihara, Kanagawa 229, Japan; \\
kawada@koala.astro.isas.ac.jp}

\and

\author{James. J. Bock}
\affil{Observational Cosmology, California Institute for Technology, 
1201 E. California Blvd., Pasadena, CA 91125; \\
jjb@astro.caltech.edu}

\altaffiltext{1}{present address: the Institute of Space and Astronautical 
Science, 3-1-1 Yoshinodai, Sagamihara, Kanagawa 229, Japan; 
masa@koala.astro.isas.ac.jp}

\begin{abstract}
We report the results of a rocket-borne observation of \c2 158\micron$\,$ line 
and far-infrared continuum emission at 152.5\micron$\,$ toward the high 
latitude molecular clouds in Ursa Major. We also present the 
results of a follow-up observation of the millimeter \co $J=1 \rightarrow 0$ 
line over a selected region observed by the rocket-borne experiment. 
We have discovered three small CO cloudlets from the follow-up \co 
observations. We show that these molecular cloudlets, as well as the MBM 
 clouds(MBM 27/28/29/30), are not gravitationally bound. Magnetic pressure and 
turbulent pressure dominate the dynamic balance of the clouds.

After removing the HI-correlated and background contributions, we find that 
the \c2 emission peak is displaced from the 152.5\micron$\,$ and CO peaks, 
while the 152.5\micron$\,$ continuum emission is spatially correlated with 
the CO emission. We interpret this behavior by 
attributing the origin of \c2 emission to  the photodissociation regions
around the molecular clouds illuminated by the local UV radiation field. 
 We also find that the ratio of the molecular 
hydrogen column density to velocity-integrated CO intensity is 
$\rm 1.19\pm0.29\times10^{20} \, cm^{-2} (K \, km\,s^{-1})^{-1}$ from the FIR
continuum and the CO data.
The average \c2 /FIR intensity ratio over the MBM clouds 
is 0.0071, which is close to the all sky average of 0.0082 reported by the 
FIRAS on the COBE satellite. 
The average \c2/CO ratio over the same regions is 420, which is significantly
lower than that of molecular clouds in the Galactic plane. 
\end{abstract}


\section{Introduction}

The \c2 ($\,^{2}P_{3/2} \rightarrow ^{2}P_{1/2}$) 157.7\micron$\,$ line 
emission is thought to be an important coolant of the diffuse interstellar 
medium(\cite{DM72}; \cite{Spi78}). 
In recent theoretical studies the \c2 line dominates the cooling  of the cold
 neutral medium(\cite{Wol95}) , and low density 
photodissociation regions(PDRs) around the surface of molecular 
clouds(\cite{HTT91}). 
The PDR models (\cite{TH85}; Hollenbach et al. 1991) are in agreement 
with observations of \c2 emission 
from individual star forming regions at the edge of the molecular clouds 
and the diffuse \c2 emission from the Galactic 
plane(\cite{Shi91}; \cite{Wri91}). 
Observation of \c2 emission from PDRs, however, has been limited to regions 
with $G_{0} > 10$, where \g0 is the far-ultraviolet(FUV) field strength 
normalized to Habing's (1968) 
estimate of the local interstellar flux, due to the low brightness of \c2 
emission from PDRs 
with \g0 as small as that found locally. Observation of 
local high latitude molecular clouds thus provides a unique example for 
study of such PDRs. The high latitude molecular clouds studied in this paper
 are believed to be located in the solar neighborhood(\cite{MBM85}, hereafter
MBM).  

Observation of diffuse \c2 line emission at high Galactic latitude is, 
however, only possible from space due to the low brightness of the line from
these regions. Balloon-borne and air-borne observations do not 
have enough sensitivity because of the enormous background from the 
earth's atmosphere and from the ambient temperature telescope.  The FIRAS 
onboard the COBE
satellite has observed the \c2 line over almost the entire sky with a 
7\arcdeg beam(\cite{Wri91}; \cite{Ben94}), which is however too large to 
observe \c2 emission from  high latitude molecular clouds.

In February 1992 we made a rocket-borne observation of \c2 158\micron$\,$ 
line and far-infrared(FIR) continuum emission of a region at high Galactic
latitude in Ursa Major with higher sensitivity  
and higher angular resolution(0.6\arcdeg) than those of the COBE/FIRAS. 
Details of the experiment are described in Matsuhara et al. (1994).
The correlation of the \c2 line with atomic hydrogen(HI) column density
in regions without appreciable CO emission 
has already been reported by Bock et al. (1993). The FIR 
continuum observations of both atomic and molecular regions have been 
presented by Kawada et al. (1994).

The main purpose of this paper is to present the first observation of the 
\c2 line from high latitude molecular clouds. We also 
report the results of a follow-up observation of the millimeter 
\co J=1$\rightarrow$0 line with a 4m 
millimeter telescope at Nagoya university.
The rocket-borne observations pass over high latitude 
molecular clouds studied by MBM: MBM 27/28/29/30. \co emission from
these high latitude clouds(hereafter MBM clouds) has already been well studied 
 by de Vries, Heithausen, \& Thaddeus (1987, hereafter VHT). 
In previous works(\cite{Boc93}; \cite{Kaw94}) we implicitly assumed there 
are no molecular regions except for those reported by VHT.  
 The follow-up CO observation enables us to check
the validity of this assumption. In this paper we assume the distance to 
the Ursa Major clouds is $D$=100pc, as adopted by VHT. 
Penprase (1993) constrained the distance of the clouds by 
optical absorption line observations toward bright stars with known distances: 
100pc$<D<$120pc.

In the following we describe the rocket-borne observation and the 
results(\S 2), the millimeter CO observation and the results(\S 3), and 
then discuss  the physical properties of the molecular clouds and the 
origin of the \c2 emission from the MBM clouds(\S 4).  

\section{Rocket-borne Observation and Results}
\subsection{Observation}

The rocket-borne instrument consists of an absolute 2-channel spectrophotometer
at the focus of 10cm liquid-helium cooled telescope which measures \c2 
157.7\micron$\,$ velocity-integrated flux(LC channel) and nearby continuum 
flux at 152.5\micron(CC channel) simultaneously. Specifications of the 
spectrophotometer are summarized in Table 1
\footnote{As a result of careful analysis of the spectral response of the 
spectrophotometer, the effective band center wavelength and the equivalent 
square band width listed in Table 1 have been slightly corrected 
from those in Table 2 of Matsuhara et al.(1994) and in Table 1 of Kawada et 
al.(1994).
The differences are very small and do not affect any scientific results of 
previously published papers.}.
Details of the instrument and the method of the \c2 observation may be found
in Matsuhara et al. (1994) and Bock et al. (1993), respectively.

\placetable{tbl-1}

The instrument was integrated with the sounding rocket S-520-15 of the 
institute of Space and Astronautical Science(ISAS) in Japan and 
launched on 1992 February 2 from the Kagoshima Space Center 
of the ISAS. Observations began at 130s after launch when a lid covering 
the telescope was opened, and ended at 480s when the instrument was
separated from the rocket payload.

Observed regions are shown in Figure~\ref{Fig1}.  For more than half of 
the observation
time the telescope was pointed at the HI hole, $(l, b) \sim 
(151\arcdeg,\,52\arcdeg)$. From 220s to 310s after launch, the telescope 
scanned along a triangular path as shown in Figure~\ref{Fig1}, through the 
high latitude molecular clouds(MBM clouds) and over the FIR bright galaxy 
M82, which was observed for in-flight calibration of the instrument.

\placefigure{Fig1}

\subsection{Results}

The data used in this paper are the LC and CC data obtained between 
222 and 357s after launch, during which the triangular scanning observation
and a part of the second pointed observation of the HI hole were made. 
 The laboratory calibration of the sensitivity, the 
in-flight subtraction of the instrumental offset, and the in-flight response 
calibration are described in Bock et al. (1993) and Kawada et al. (1994). 
The response of the spectrophotometer was monitored by an 
internal calibration lamp in flight, and was closely matched with the response
observed in the laboratory.   We 
estimate the uncertainty of the absolute brightness calibration is 6.4\% for 
152.5\micron$\,$ continuum, and 8.4\% for the \c2 line.  

As described in Bock et al., the \c2 line intensity is extracted from the 
difference of LC signal and CC signal. Thus the derived \c2 intensity depends 
on the spectral index $s$ ($I_{\nu} \propto \nu^{s}$) of the continuum between
152 and 158\micron, and 
we previously assumed  $s=-1$. In Kawada et al. (1994), we reported the 
FIR continuum spectra of the observed regions using the broad-band 
photometer data. We derive $s=-0.25$ for the spectrum at the HI hole
and $s=-0.65$ for the spectrum of a bright IR cirrus. We find that 
in the worst case the 
assumption $s=-1$ leads to 10\% systematic error in the \c2 intensity, 
so in this paper we use $s=-0.5$. The statistical error is derived from the 
data obtained during the pointed observation of the HI hole between 310s 
and 357s after launch as in Kawada et al.: 
$\rm 0.7pW cm^{-2}sr^{-1}$ for CC, $\rm 0.8pW cm^{-2}sr^{-1}$ for LC, 
respectively. The 3$\sigma$ [CII] line detection limit is
estimated to be $\rm 2.7\times10^{-7}\, ergs \, s^{-1}cm^{-2}sr^{-1}$.

We compare the resulting \c2 and 152.5\micron$\,$ continuum data with the HI 
data(Heiles 1992, private communication) and the CO data of VHT. 
After Bock et al. was published, we determined the positions of 
the telescope more accurately($\pm 3\arcmin$) using the near-infrared
spectrometer data(\cite{syu94}; \cite{Mrm94}). We find that the assumed 
pointing data used in Bock et al.  were systematically offset by 
$\sim$8\arcmin, so we re-sampled the HI data and the CO data as we did in
Kawada et al. The pointing correction is significant for the CO data, since
the size of the CO clouds is generally smaller than the 36\arcmin$\>$ beam. 

The molecular clouds coincide with HI clouds emitting comparable
\c2 line and FIR continuum. Hence in order to determine the \c2 and FIR 
continuum intensities emitted from the molecular clouds alone, we have to 
estimate the contributions from the HI and background components to 
the observed \c2 and continuum emission.
As shown in \S 3.2, appreciable CO emission is found only for beams 
with $N_{\rm HI} > 2\times10^{20} \rm cm^{-2}$. Thus we 
assume that all data with $N_{\rm HI} < 2\times10^{20} \rm cm^{-2}$
do not have appreciable CO emission, and bin these $N_{\rm HI}$ data in 
$0.1\sim0.2\times 10^{20} \rm cm^{-2}$ intervals. Several 
beams(1, 2, 4, 5, 6, 7, 15, 16) are also denoted which have $N_{\rm HI} 
> 2\times10^{20} \rm cm^{-2}$ but lack appreciable CO emission.  The 
results are shown in Figure~\ref{Fig2}.

\placefigure{Fig2}

\section{Millimeter \co Observation and Results}
\subsection{Observation}

We use the millimeter-wave \co emission data to distinguish regions with
significant molecular hydrogen.  However, the beam positions of the 
rocket-borne instrument are not fully mapped by VHT. 
Thus, in order to increase the number of beams with \co data, we have 
observed at eight beam positions along the rocket observation 
path(labeled in Figure~\ref{Fig1}) with the Nagoya University 4m millimeter 
telescope(\cite{fus92}) in 1994 February -- March. Among them, beams 7, 8 
and 9 were also mapped by VHT.

Because the beam of the rocket observation(36\arcmin$\,$ in diameter) is much
larger than that of millimeter CO observation(2\arcmin.7),
we mapped at each beam position over a $40\arcmin \times 40\arcmin$ or wider 
area. The mapping grid sizes were 4\arcmin -- 5\arcmin.66, except for 
parts of beams 5, 8, 9 where a further 5$\times$5 map was made around the 
emission peak with a 2\arcmin$\,$ grid. 
Integration time per each grid point was 3 -- 6min with typical rms noise of 
0.3K per spectral channel($\rm 0.1km\,s^{-1}$).
We used a frequency switching method to obtain the line signal 
without observing a spatial reference point. Typically data at  
$v < -4 \rm km\,s^{-1}$  
and $v > 10 \rm km\,s^{-1}$ are used as the baseline for the \co line spectrum.
We frequently observed Orion A for brightness calibration, and assumed 
its brightness temperature is 60K. We also frequently observed the peak 
brightness position in beam 9 to correct the effect of varying airmass, 
which we found to be negligible. 

\subsection{Results}

We obtained a positive CO detection 
in 5 beams(3, 4, 5, 8, and 9), all of which have relatively large
$N_{\rm HI}$($>2\times10^{20}\rm cm^{-2}$).   
Contour maps of the velocity-integrated 
brightness temperature of the 
CO line($W_{\rm CO}$, in $\rm K\,km\,s^{-1}$) of these beams are shown in 
Figure~\ref{Fig3}, superposed on 100\micron$\,$ maps reproduced from the IRAS 
Sky Survey Atlas(ISSA) by reducing the intensity by a factor of 0.72, 
following the COBE/DIRBE calibration (\cite{Whe94}). The spatial 
distribution of the integrated CO line 
in beams 8 and 9(Figs.~\ref{Fig3}d and \ref{Fig3}e) are in good agreement with 
those of VHT.  However, we find that the absolute brightness temperature of 
VHT is systematically lower by a factor of 0.68.
Since the purpose of the CO observation in this work is to supplement the 
observation of VHT, we scaled the brightness temperature of all data so that
our  data of beams 8 and 9  match the data of VHT.

\placefigure{Fig3}
\placefigure{Fig4}
\placetable{tbl-2}

We discovered a few isolated, small CO cloudlets in beams 3, 4, 5. The 
typical spectra of the CO cloudlets are shown in Figure~\ref{Fig4}. 
The spectral data are integrated in $\rm 1km\,s^{-1}$ bins and fitted by a 
gaussian in order to evaluate the line widths. Location, size, line width,
and the derived physical properties(details are described in \S 4.3) of the 
cloudlets are listed in Table 2. 

In order to compare with the rocket-borne data, average $W_{\rm CO}$ was 
calculated by convolving the CO data with the 36\arcmin$\>$ beam of the 
rocket-borne instrument. Although we have a
positive CO detection in beams 4 and 5, these cloudlets are very small 
compared with the rocket beam and thus when averaged over the 
36\arcmin$\>$ beam do not give appreciable average $W_{\rm CO}$(more than 
$3\sigma$: $\rm \sim 0.2K\,km\,s^{-1}$ ) for these beams. In the averaging 
procedure the error is dominated by the poor quality of the spectral 
baselines incurred at some grid positions due to temporal changes 
of the weather condition during the observation. Only for beams 3, 8, 
and 9 did we obtain appreciable $W_{\rm CO}$($\rm > 0.2K\,km\,s^{-1}$).

\section{Discussion}
\subsection{HI regions}

In previously published papers(\cite{Boc93}; \cite{Kaw94}), we 
 assumed that there are no molecular regions except for those mapped by VHT.
 Outside the regions mapped by VHT, only beam 3 has appreciable CO emission. 
Therefore, the previous results  on the physical properties 
of the HI regions do not change significantly. We summarize 
the results for the HI regions for purposes of the following analysis of 
the molecular regions.

The correlation between 152.5\micron$\,$ continuum emission and HI column 
density is shown in Figure~\ref{Fig2}a, and the correlation between \c2 
emission and HI column density is shown in Figure~\ref{Fig2}b. The excellent
correlation of the 152.5\micron$\,$ continuum intensity and HI column density
strongly suggests that instrumental effects such as detector hysterisis are
small. Laboratory data demonstrate that detector hysterisis is still further
removed in the \c2 profile which is derived by difference of signals between
two well-matched detectors. 

The 152.5\micron$\,$ continuum emissivity per HI column density is 
determined to be $\lambda I_{\lambda} / N_{\rm HI} = 3.62\pm0.38 \times
10^{-32}\,\rm W\,sr^{-1}$, and the \c2
cooling rate per atomic hydrogen is determined to be $\Lambda_{\rm C II}
\equiv 4 \pi I$(\ion{C}{2})$ / N_{\rm HI} = 
1.32\pm0.24 \times 10^{-26} \rm ergs\,s^{-1}\,H_{atom}^{-1}$. This value is
in good agreement with the average cooling rate at high latitude obtained
 by recent analysis of the COBE/FIRAS data: $\Lambda_{\rm C II} = 
1.45 \times 10^{-26} \rm ergs\,s^{-1}\,H_{atom}^{-1}$(\cite{Dwe97}).
If only data with $N_{\rm HI} < 2 \times 
10^{20}\rm cm^{-2}$ are considered, these values change  little:
 $\lambda I_{\lambda} / N_{\rm HI} = 2.66\pm0.45 \times 10^{-32} \rm 
W\,sr^{-1}$, $\Lambda_{\rm C II} = 1.64\pm0.39 \times 10^{-26} \rm 
ergs\,s^{-1}\,H_{atom}^{-1}$.  As already noted by Bock et al.,
inclusion of the region with relatively low line-to-continuum ratio
(beams 4, 5, 6, and 7) tends to increase the FIR 
continuum emissivity and to decrease the \c2 line cooling rate. 
These uncertainties in $\lambda I_{\lambda} / N_{\rm HI}$ and 
$\Lambda_{\rm C II}$ of the HI correlated components are, however, 
not significant 
in deriving the intensity of \c2 line and FIR continuum from the molecular 
regions, and thus do not appreciably affect the following discussion. 

\subsection{The Conversion Factor $\bf X=N(H_{2})/W_{CO}$}

We subtract the HI and background contributions from the 152.5\micron$\,$
continuum  and obtain the excess 
FIR emission plotted against $W_{\rm CO}$ in Figure~\ref{Fig5}. 
The fitted line in Figure~\ref{Fig5} gives 
$\lambda I_{\lambda}/W_{\rm CO} = 8.6\pm2.1\,\rm  pW\> 
cm^{-2}sr^{-1}\,(K\,km\,s^{-1})^{-1}$ with no significant 
residual FIR emission at $W_{\rm CO}=0$.
The conversion factor $X \equiv N({\rm H_2})/W_{\rm CO}$, the parameter 
used to estimate the molecular column density over CO clouds, is derived
 assuming constant FIR emissivity per hydrogen nucleus in both the 
molecular and the HI regions. We find 
$ X = 1.19\pm0.29\times10^{20}\,\rm cm^{-2}\, (K\,km\,s^{-1})^{-1}$, 
which is not significantly different from the value derived by Kawada 
et al. (1994). 

\placefigure{Fig5}

The conversion factor X is presumably not constant, varying significantly from 
region to region. VHT gives $X = 0.5\pm0.3
\times10^{20}\,\rm cm^{-2}\,(K\,km\,s^{-1})^{-1}$ over a 
larger region of the molecular clouds using IRAS 100\micron$\,$ data. 
We independently examine the factor X on small scales by studying the 
correlation between CO and IRAS 100\micron$\,$ emission for beams 3, 8, and
9 where CO counterparts are clearly visible in the IRAS 100\micron$\,$ 
maps(Fig.~\ref{Fig3}).
 No correction for the HI component is applied to the 
IRAS 100\micron$\,$ data,  so significant spatial structure in HI column
density invalidates this analysis.  
From the slopes of the fitted lines and the global FIR spectrum of dust 
emission($\lambda I_{\lambda}(152.5 \micron) / \lambda I_{\lambda}(100
 \micron) = 1.5(+0.4, -0.3)$; Kawada et al.), we derive the factor
X for beams 3, 8, 9 as $0.54(+0.27, -0.24)$, $1.0(+0.43, -0.36)$, 
$0.86(+0.41, -0.35) \, \times10^{20}\,\rm cm^{-2}\,(K\,km\,s^{-1})^{-1}$, 
respectively. While the factor X  for beams 8 and 9 is consistent with the
global value, that of beam 3 is appreciably lower. 

\subsection{Physical Properties of the Molecular Clouds}
 
In Table 2 the mass of the clouds calculated from the integrated 
CO flux 
and the (global) conversion factor X are listed. The distance to the 
clouds is not known for the small cloudlets in beams 3, 4, and 5, and thus 
the mass of these cloudlets is calculated by assuming the same
 distance of that of the MBM clouds: $D=100$pc.
In addition, considering the local 
variation of the factor X as described in the previous subsection, the mass 
of the cloudlets may be smaller by a factor of 2. 

The average volume density of molecular hydrogen, calculated by assuming a 
spherical cloud of radius $R$ ( $=(A/\pi)^{1/2}\ast D$, where $A$ is solid 
angle of the cloud), is also shown in Table 2. 
 For the MBM clouds, 
the assumption of a spherical geometry is clearly not applicable. However,   
the average density derived by dividing the column density of  molecular
 hydrogen by the width of the CO filament($\sim$1pc in the plane of sky) 
ranges from 10 -- $50\rm cm^{-3}$, similar value to that calculated by
assuming a spherical cloud.
It is interesting that the density of the small cloudlets in beams 3, 4, and 5
is much larger than that of the MBM clouds. This may suggest that the MBM 
clouds are very clumpy, consisting of smaller, denser cloudlets like those 
in beams 3, 4, and 5. Another interesting fact is that the average number 
density of these clouds is too low for CO molecules to survive 
photodissociation rate from the interstellar UV field. For example, 
$n({\rm H_2})R \sim 2\times10^{20}\,\rm cm^{2}$, corresponding to 
$A_{V} \simeq 0.1\rm mag$, for CO 145.55+43.32(the cloudlet in beam
3). Thus, to survive the photodissociation, the clouds may consist
of many small and dense clumps. In this case, since we did not observe such 
clumpy structure with the 4m telescope, each clump must be  small enough to
 be unresolved($<2\arcmin$). Alternatively, 
the UV photons may be strongly attenuated by intervening clouds between the UV 
sources and the CO cloudlets(see \S 4.4).

In Table 2 rough estimates of the turbulent pressure($P_{turb}$) 
and the gravitational pressure($P_{grav}$) are listed for the cloudlets in 
beams 3, 4, and 5, and for the MBM clouds.
The turbulent pressure of the clouds is estimated from the \co velocity 
width($\Delta v_{\rm FWHM}$; Fig.~\ref{Fig4}), the calculated mass $M$ 
of the cloud, and the effective radius  $R$, using 
$4 \pi R^{2} P_{turb} = 3 M \Delta v_{\rm FWHM}^{2} / (8 \ln 2) R$.
The gravitational pressure,  
calculated as $ 4 \pi R^{2} P_{grav} = 3 GM^{2}/ 5 R^{2}$, 
is orders of magnitude smaller than the turbulent pressure. 
Since the external thermal pressure of the HI gas(estimated by the \c2 line 
cooling rate, see Bock et al. (1993)), $\sim 10^{-13}\,\rm ergs\,cm^{-3}$, 
is also much smaller than the turbulent pressure, the cloud cannot be 
bound by gravitation or by external thermal pressure. Joncas et al. (1992) 
calculated the magnetic pressure, 
by using the magnetic field strength at $(l, b)=(142\arcdeg.6, 38\arcdeg.4)$
(close to beam 9) measured by Heiles (1989), to be $1\times10^{-11}\,\rm 
ergs\, cm^{-3}$. Joncas et al. also estimated the turbulent pressure of 
the HI gas to be $4\times10^{-11}\,\rm ergs\,cm^{-3}$ from the velocity width 
of the HI emission line.
Thus as Joncas et al. concluded for the HI clouds, the turbulent pressure in 
these molecular clouds is possibly balanced by the magnetic pressure as well 
as the external turbulent pressure of the HI gas.

\subsection{\c2 Emission from the MBM Clouds}
 
In Figure~\ref{Fig6} we compare the spatial distribution of  $W_{\rm CO}$ 
and excess({\it i.e.} corrected for HI and background contributions)
 \c2 line emission. 
Interestingly, the spatial distribution of the excess \c2 emission is very
much different from that of $W_{\rm CO}$ while the 152.5\micron$\,$ continuum
emission is correlated with $W_{\rm CO}$(Fig.~\ref{Fig5}). The \c2 emission 
peak is displaced from the CO peak toward lower latitude by $\sim$1\arcdeg$\,$.
In the following we discuss the origin of the excess \c2 emission
and present the most plausible interpretation of this interesting behavior.

\placefigure{Fig6}

 Excess \c2 emission could possibly originate from \ion{H}{2} gas ionized 
by Lyman
continuum photons or Galactic shock waves. Such ionized component of the local
interstellar medium, the warm ionized medium(WIM), is responsible for diffuse 
$\rm H\alpha$ emission(\cite{Rey93}, and references therein). 
However, the $\rm H\alpha$ intensity over the entire observed region  is unusually 
low($I(\rm H\alpha) \simeq 0.7 \pm 0.5 \times
10^{-7}  \,\rm ergs\,s^{-1}cm^{-2}sr^{-1}$; Reynolds, private communication)
compared with the average $\rm H\alpha$ intensity for $\vert b \vert \leq 
15\arcdeg$,
$<I(\rm H\alpha)> \> \simeq 2.9\times10^{-7} \csc \vert b \vert \,\rm 
ergs\,s^{-1}cm^{-2} sr^{-1}$(eq. (8) of \cite{Rey92}). In the low-density 
limit the \c2 line intensity is proportional to the emission measure 
and hence proportional to the $\rm H\alpha$ intensity, $I$(\ion{C}{2})$\simeq 
1.45 I(\rm H\alpha)$ for $T=10^{4}$K and 
$\rm C^{+}/H=3.3\times10^{-4}$(eq.(6) of \cite{Rey92}). Thus we conclude that
\c2 emission associated with the WIM  is negligible,
$I$(\ion{C}{2})$ \leq 2\times10^{-7}\,\rm ergs\,s^{-1}cm^{-2}sr^{-1}$. 

The excess \c2 emission may originate from shock-compressed HI gas  since 
the MBM clouds are located in 
the north celestial loop, a large($\sim20\arcdeg$) loop of HI gas produced 
by stellar winds or supernovae explosions(\cite{HH74}). 
Since \c2 emission is a collisionally excited line, emission from the 
shock-compressed(or shock-heated) HI gas is stronger  than that of ambient 
low-density HI gas.
 In order to explain the observed \c2 
strength of the beams 10 -- 14, for example, a few times 
higher density than that of the ambient HI gas is required. We think  this
is quite unlikely because beams 8, 9, 15, and 16, located on local maximum
of the HI loop structure, show less \c2 
emission than that from beams 10 -- 14. 

We propose that the excess \c2 emission originates from  PDRs formed around 
molecular clouds, where  CO molecules are photodissociated by FUV 
photons to produce $\rm C^{+}$ ions(and $\rm C^0$) while most of the 
hydrogen remains in molecular form(\cite{vDB88}; Hollenbach et al. 1991).
At the surface of a PDR, \c2 158\micron$\,$ line emission dominates 
the gas cooling, while grain photoelectric heating is dominant over almost the
entire PDR. Therefore, the spatial variation of \c2 emission 
should match the variation of the grain photoelectric heating. According to 
models of Hollenbach et al. and Wolfire et al. (1995), 
the efficiency of grain photoelectric heating is almost constant in the case 
of $G_{0}\sim 1$, $n({\rm H_2})\sim20\rm cm^{-3}$(average density of the 
MBM clouds, Table 2), and $T\sim100$K. Thus the variation of 
the grain photoelectric heating rate must be due to variation of the 
dust-to-gas mass ratio or  variation of the FUV field strength. Since the 
FIR continuum emission spatially well-correlated with the CO emission, the 
variation of dust-to-gas mass ratio is unlikely, and variation of 
the FUV strength is most probable.

The variation of FUV strength can be explained by the location of a FUV 
source and the distribution of attenuating material. Because the MBM
clouds are located at about 60pc above the Galactic plane(calculated from 
$D=100$pc and $b\simeq38\arcdeg$), cloud surfaces facing 
the Galactic plane are presumably illuminated  by stronger FUV photons. 
Using the average density $n({\rm H_2})=20\rm cm^{-3}$ of the MBM clouds and 
the length of the molecular 
filament on the sky($\sim$10pc), the optical depth through the excess \c2 
emission region is estimated to be $A_{\rm V}=0.65 \rm mag$ or 
$A_{\rm FUV}=1.2 \rm mag$( $A_{\rm FUV}/ A_{\rm V}=1.8$; \cite{DDB80}). 
The HI gas further contributes 
to the attenuation. Thus, the FUV photons  are significantly attenuated toward 
higher Galactic latitudes by the molecular and the HI clouds, resulting in 
weaker \c2 emission. 
This might be the reason why the small CO  cloudlets found at higher 
latitude regions(in \S 4.3) withstand photodissociation and indeed the \c2
line to FIR continuum ratio is anomalously low. 
The depth of the \c2 emission region agrees well with the theoretical 
thickness of the $\rm C^{+}$ region formed on the surface of a molecular cloud 
illuminated by the standard interstellar FUV 
flux($A_{\rm V}(\rm C^{+})\sim 0.5 mag$; \cite{vDB88}), supporting 
this interpretation.  
 
\subsection{Overall Observational Properties of the MBM clouds}

The integrated FIR continuum intensity may be calculated from the observed 
152.5\micron$\,$ intensity assuming the FIR spectrum determined by the other 
photometric channels onboard 
the instrument($T=16.6$K gray-body with an emissivity $\propto \lambda^{-2}$ ; 
\cite{Kaw94}).
By averaging the \c2 intensity and the integrated FIR continuum intensity 
over the CO regions(beams 8 -- 14), we obtain an average ratio of
\c2/FIR$\simeq0.0071$, which is close
to the all sky average of $0.0082$ reported by the FIRAS(\cite{Wri91}) which
is heavily weighted toward the Galactic plane. Using the FIR continuum 
emissivity obtained for the HI clouds(\S 4.1), the observed \c2/FIR ratio
corresponds to a [CII] cooling rate of $\Lambda_{\rm C II} \simeq 3 
\times 10^{-26} \,\rm ergs \, s^{-1} H_{nuc}^{-1}$, which is a factor of 
two larger than that of the HI clouds. This may indicate that the heating 
efficiency is higher 
in the PDRs around the molecular clouds than in the diffuse HI clouds.
We also obtain an average ratio of \c2/CO$\simeq420$ by taking the observed 
$W_{\rm CO}$ and converting to brightness by $\rm 1K\,km\,s^{-1} = 
1.6\times10^{-9}\,ergs\, s^{-1}cm^{-2}sr^{-1}$. This ratio is significantly 
smaller than that
of molecular clouds in the Galactic plane(1500; \cite{Shi91}), as expected
if \g0 is lower at high Galactic latitude than on the Galactic 
plane(Hollenbach et al. 1991).

\section{Summary}

We report the first \c2 158\micron$\,$ line observation of high latitude 
molecular clouds. The \c2 line and 152.5\micron$\,$ continuum data  
 are obtained using a rocket-borne spectrophotometer which 
observed a region at high Galactic latitude in Ursa Major including 
the MBM  clouds. We also present the results of a follow-up 
observation of the millimeter \co $J=1\rightarrow0$ line. 

Over the observed region  the ratio of molecular 
hydrogen column density to velocity-integrated CO intensity is 
$ X = 1.19\pm0.29\times10^{20}\,\rm cm^{-2}\, (K\,km\,s^{-1})^{-1}$, 
assuming a constant FIR emissivity per 
hydrogen nucleus for both the molecular and the atomic regions.  
We have also discovered three small CO cloudlets as a result of the
\co observation of the selected region in Ursa Major. Using the factor X and
the observed width of the CO line, we estimate the
mass and the turbulent pressure of the CO cloudlets. We find that 
these molecular cloudlets, as well as the MBM  clouds, are not 
gravitationally bound. Magnetic pressure and turbulent pressure 
dominate the dynamic balance of the clouds.

After subtracting the HI and the background contributions, we compare the 
spatial distribution of the \c2 line, the 152.5\micron$\,$ continuum, and 
the \co line intensity. The \c2 emission peak is displaced from  the 
\co line intensity peak, while the 
152.5\micron$\,$ continuum emission is spatially correlated with the \co line
emission. We discuss the origin of the \c2 emission and conclude that
the \c2 emission most probably originates in PDRs around the molecular clouds 
illuminated by the local UV radiation field.

The average ratio of the \c2 line to the integrated FIR continuum 
intensity over the MBM clouds was found to be 0.0071, which is close to 
the all sky average of 0.0082 reported by the FIRAS on the COBE. 
This \c2/FIR ratio corresponds to the [CII] cooling rate 
$\Lambda_{\rm C II} \simeq 3 \times 10^{-26} \,\rm ergs \, s^{-1} 
H_{nuc}^{-1}$, which is a factor
of two larger than that of HI clouds. The \c2/CO ratio 
calculated in the same manner was 420, which is significantly lower than that
of molecular clouds in the Galactic plane.

\acknowledgments

We would like to thank to all the collaborators of the S-520-15 rocket 
experiment: T. Matsumoto, P. L. Richards,  A. E. Lange, V. V. Hristov, S. 
Matsuura, and P. D. Mauskopf. We also thank H. Okuda and the ISAS launch 
staff for their kind support. We thank R. Reynolds for providing
the $\rm H\alpha$ data toward the Ursa Major region.
This work was supported by the ISAS, by the Daiko Science Foundation, by 
grant-in-aids (No. 03249213, 03740128, \& 04233215) from the Ministry of 
Education, Science and Culture in Japan, and by the National Aeronautics and 
Space Administration  under grants NAGW-1352 and NGT-50771.

\clearpage

\clearpage

\figcaption[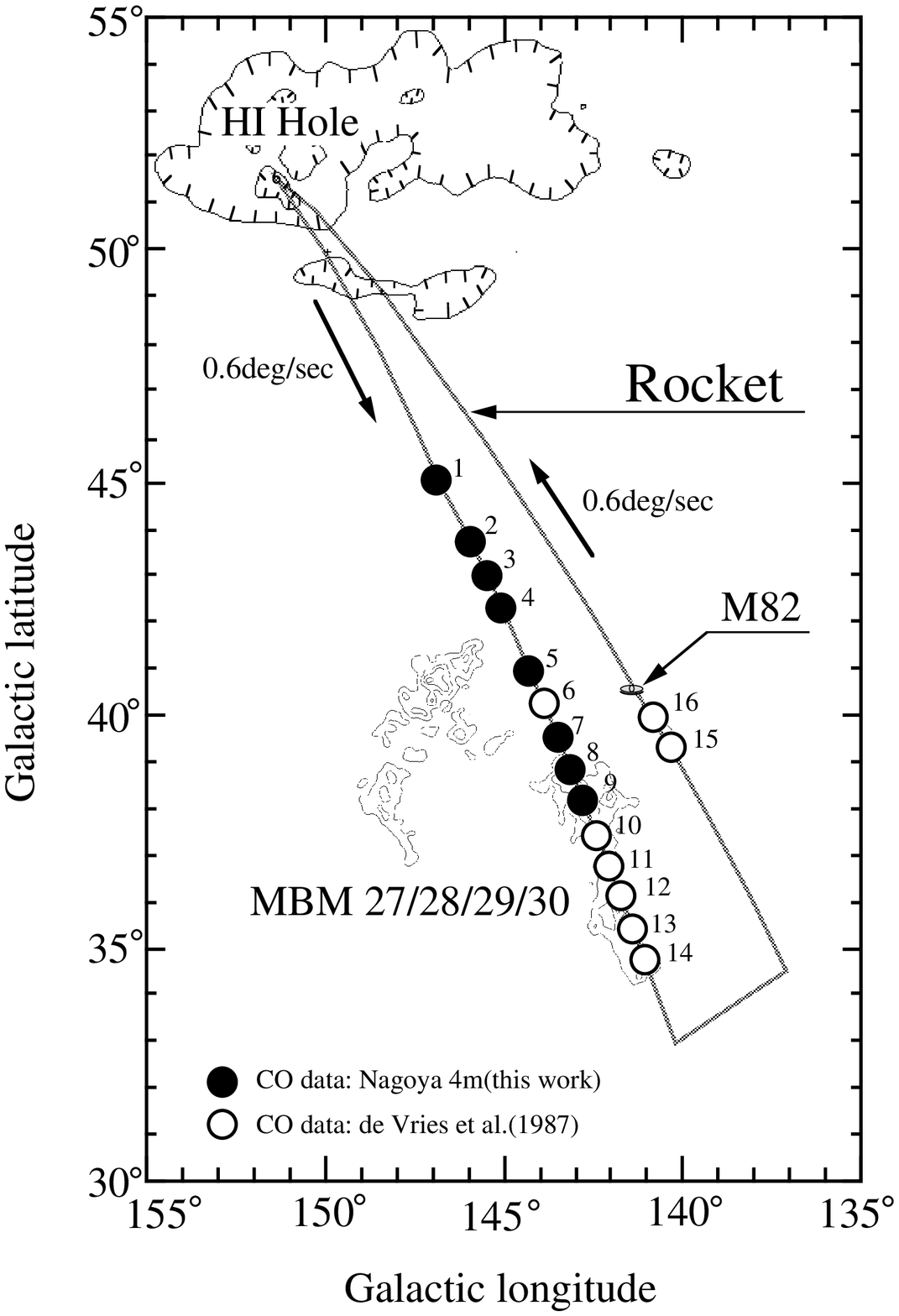]{Rocket-borne \c2 and FIR continuum 
observations were performed along a triangular path crossing the high latitude 
molecular clouds(MBM 27/28/29/30) in Ursa Major. The contour map of the MBM 
clouds is from VHT. For beams shown by the 
circles, \co $J=1\rightarrow0$ data are also available. Beams shown by filled
circles are mapped with the 4m millimeter telescope at Nagoya University. 
Size of the circles corresponds to the beam size of the rocket observation. 
\label{Fig1}}

\figcaption[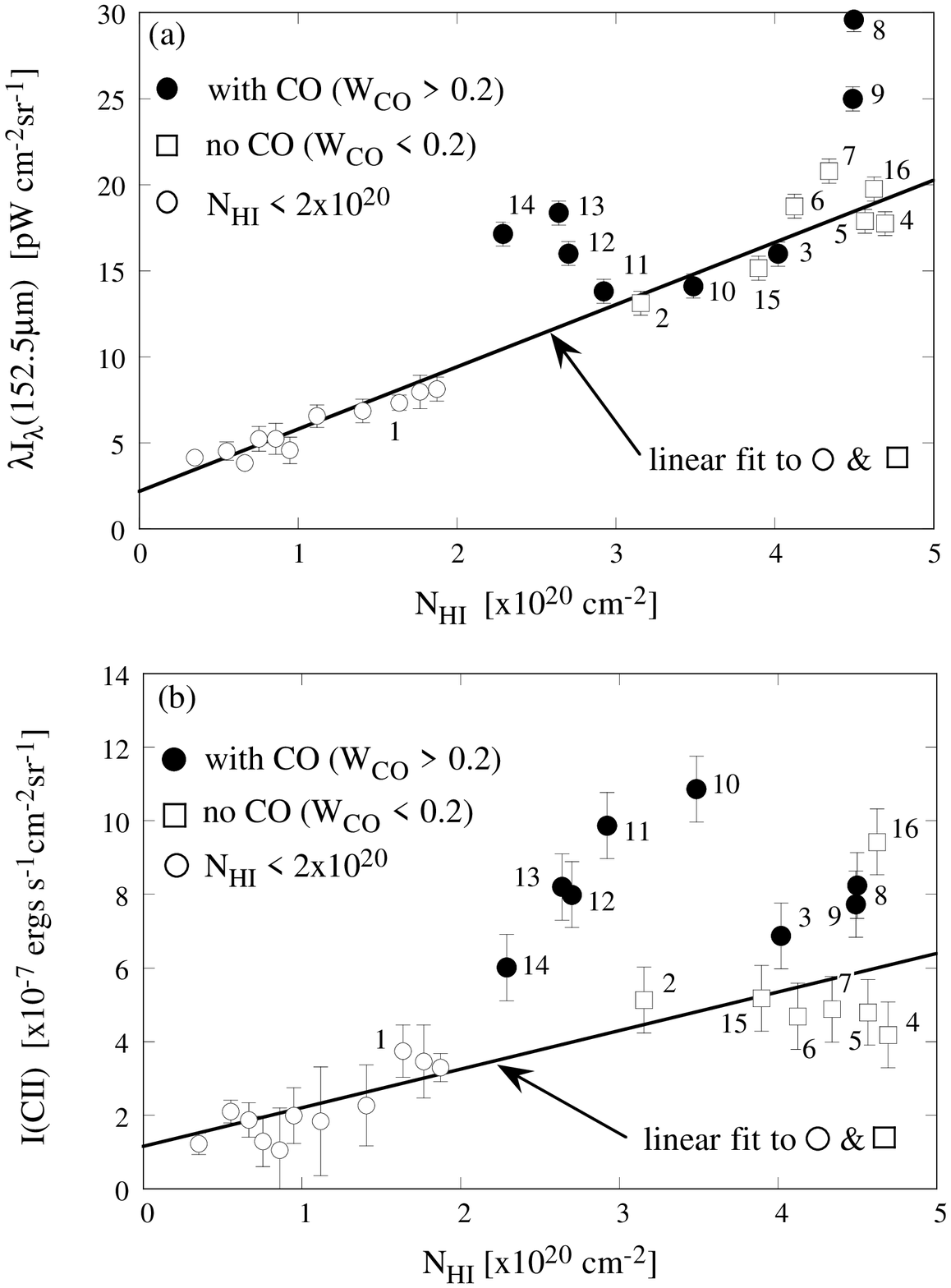]{(a) 152.5\micron$\,$ continuum intensity  vs. 
atomic hydrogen column density($N_{\rm HI}$), and  (b) \c2 
158\micron$\,$ line intensity vs. $N_{\rm HI}$. For most of the beam positions
 with $N_{\rm HI}<2\times10^{20}\,\rm cm^{-2}$, no CO data are available.  
From the correlation over beams with
$N_{\rm HI}<2\times10^{20}\,\rm cm^{-2}\,$  and beams where we confirmed 
there is no significant CO emission(beams 2, 4, 5, 6, 7, 15, 16, where 
$W_{\rm CO} < 0.2 \rm K\,km\,s^{-1}$), we obtain the average FIR continuum 
emissivity and the \c2 cooling rate of the HI regions(solid lines). 
\label{Fig2}}

\figcaption[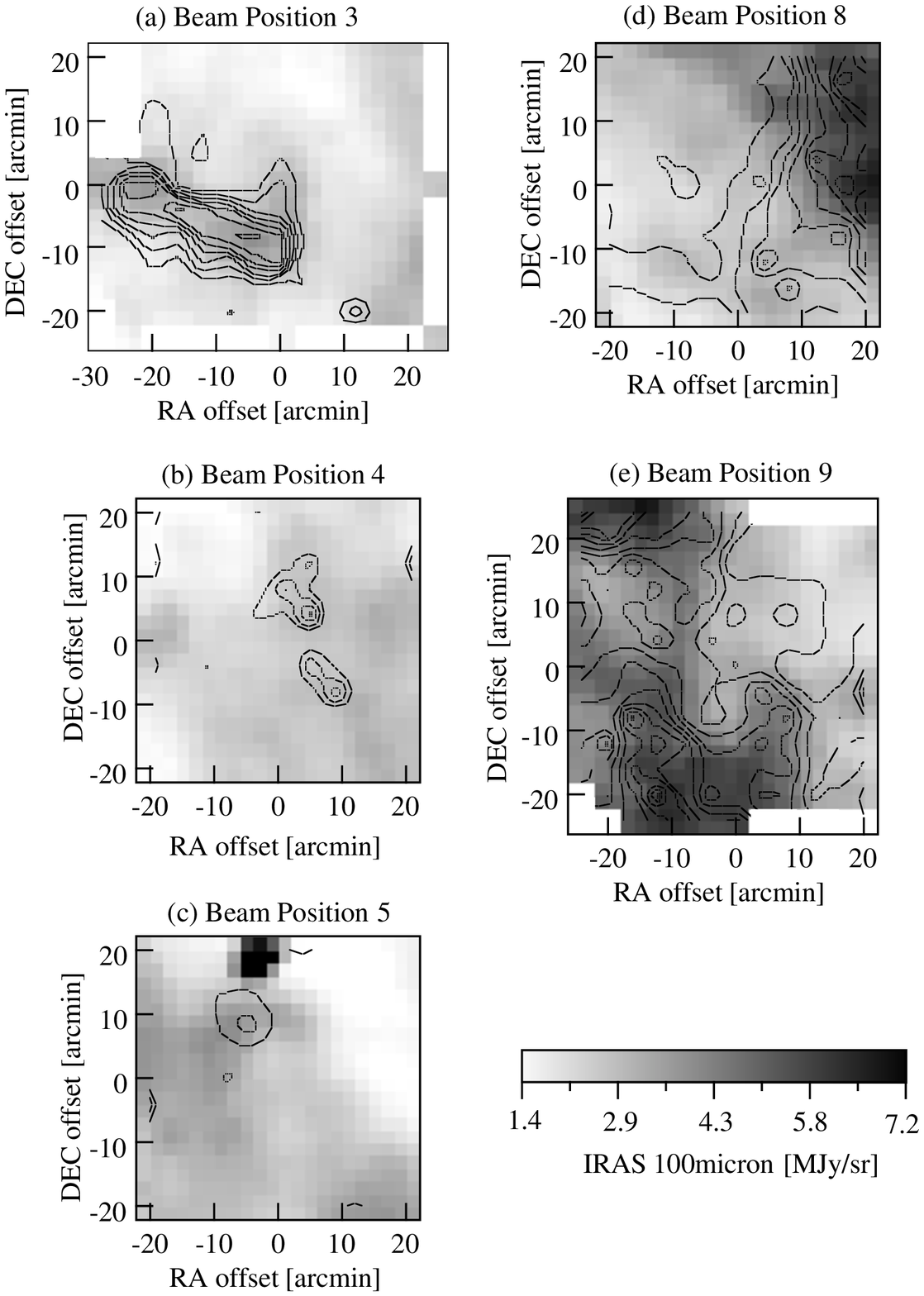]{\co $J=1\rightarrow0$ contour maps obtained with the 4m 
millimeter telescope at
Nagoya University, superposed on IRAS 100\micron$\,$ maps. Eight
beams(filled circles in Fig.~\ref{Fig1}) were surveyed for CO 
emission(typically 
$40\arcmin \times 40\arcmin$, 2\arcmin -- 8\arcmin grids) with 
positive detection for the five beams shown here. The contours are spaced at
$\Delta W_{\rm CO} = 0.6 \rm K\,km\,s^{-1}$ for beams 3, 8, 9 and 
$0.4 \rm K\,km\,s^{-1}$
 for beams 4, 5. The lowest contour level is $0.6 \rm K\,km\,s^{-1}$.
\label{Fig3}}

\figcaption[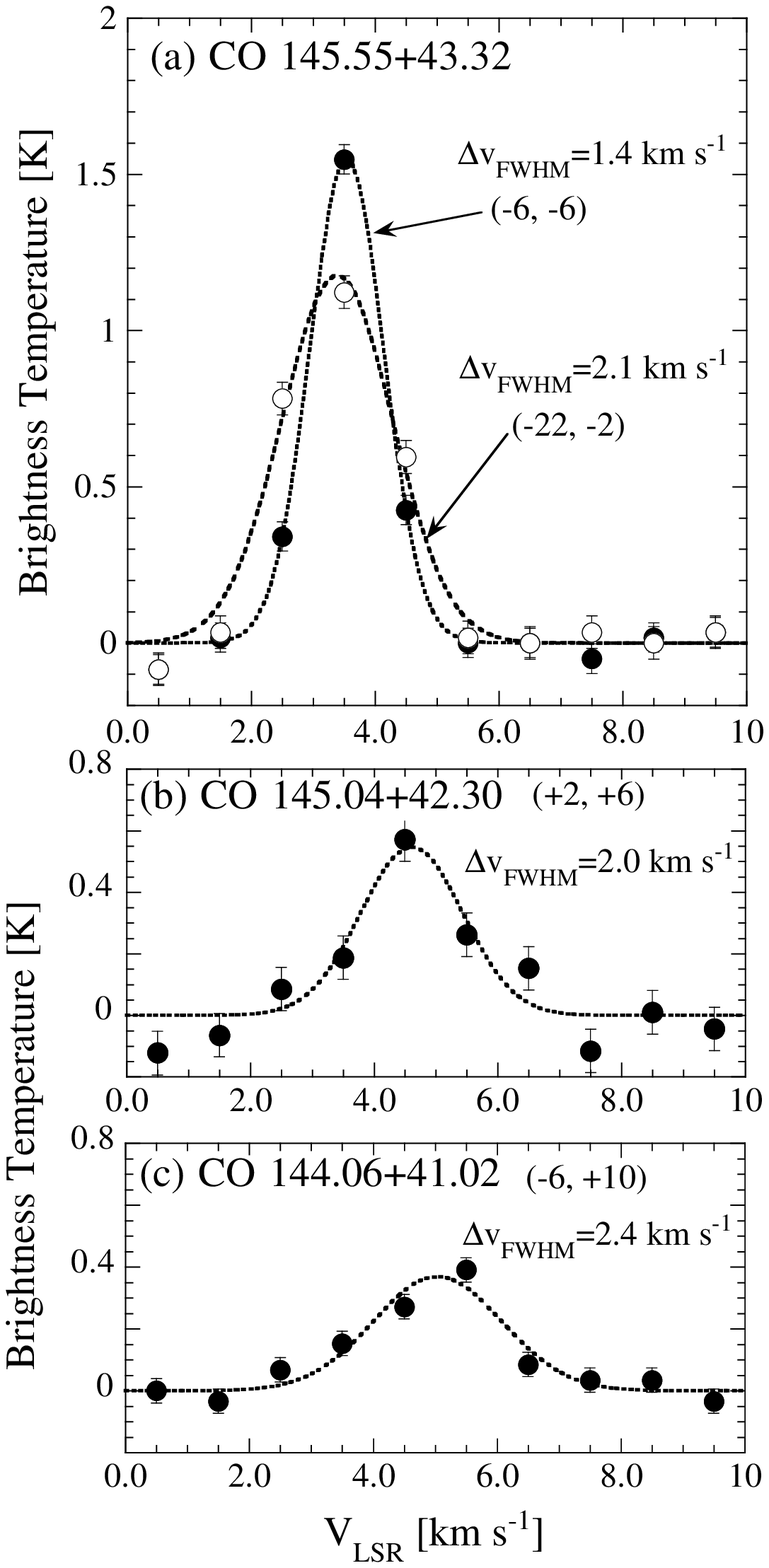]{Typical \co $J=1\rightarrow0$ spectra for beams 3, 4, 
and 5, showing the newly discovered small molecular cloudlets. (a) Spectra at 
($\Delta \alpha$, $\Delta \delta$)=(-6, -6) and (-22, -2) in 
Fig.~\ref{Fig3}a, (b) spectrum 
at (+2, +6) in Fig.~\ref{Fig3}b, (c) spectrum at (-2, +10) in 
Fig.~\ref{Fig3}c. 
\label{Fig4}}

\figcaption[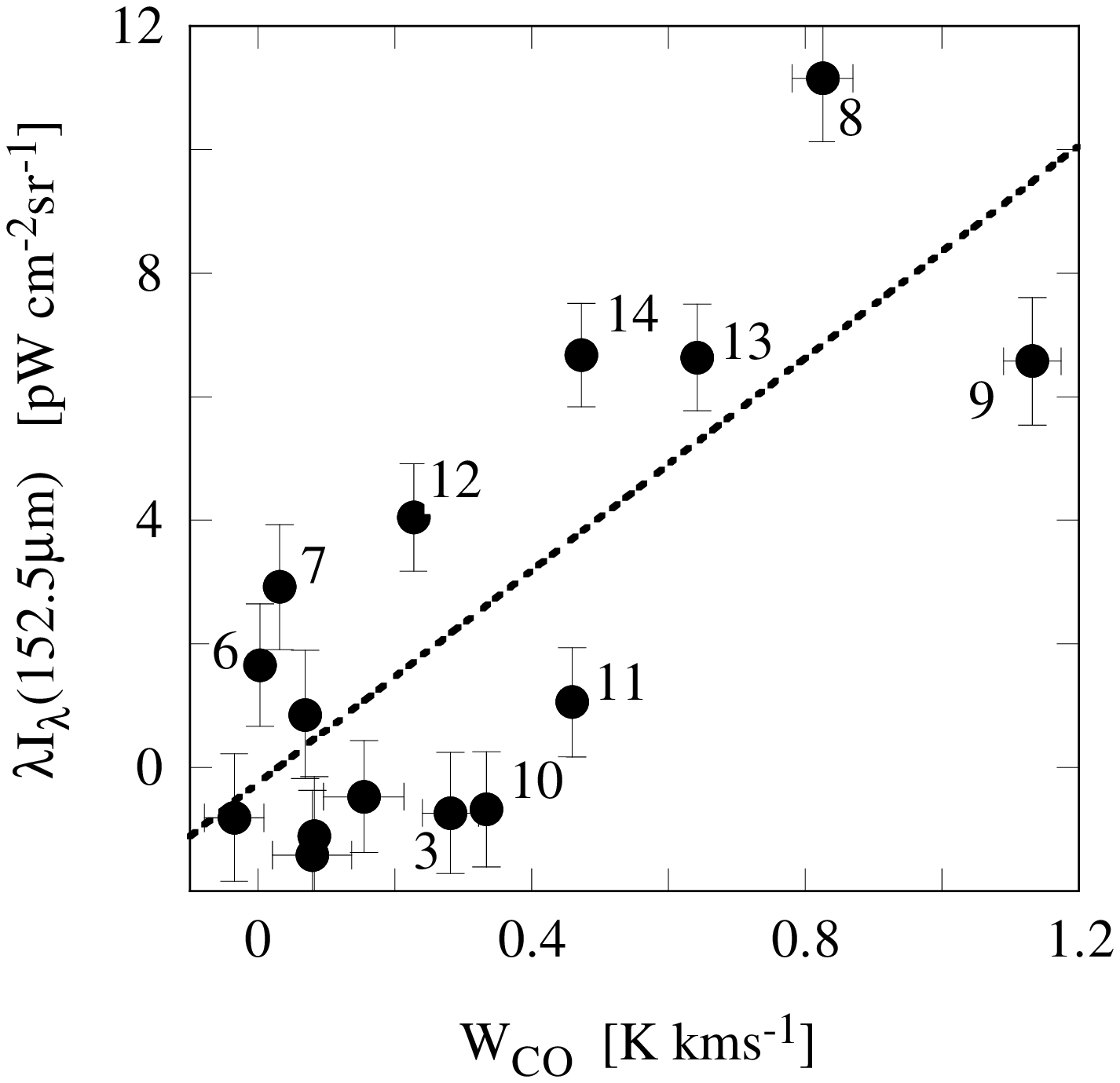]{Excess(corrected for HI and background contributions) 
152.5\micron$\,$ continuum emission vs. $W_{\rm CO}$. The best linear fit 
 is shown by the dashed line. \label{Fig5}}

\figcaption[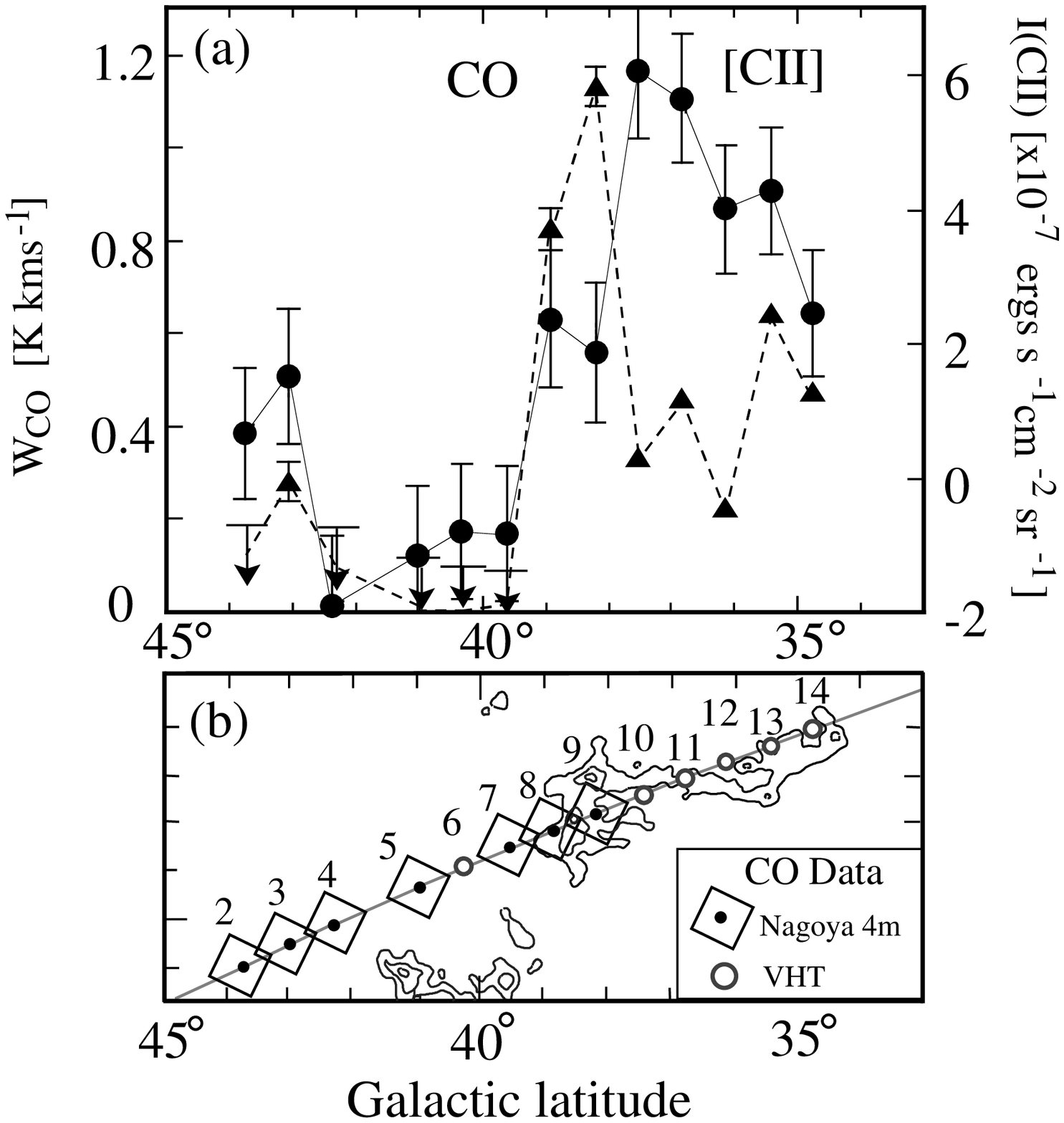]{(a) Spatial distribution of $W_{\rm CO}$(filled 
triangles) and excess \c2 line intensity(filled circles) for beams 2 -- 14
 shown in (b) a part of the observed region.  The upper
limits of $W_{\rm CO}$ are  $3\sigma$. \label{Fig6}}

%
\end{document}